\documentclass[twocolumn]{aastex702}

\graphicspath{{./}{Figures/}}

\usepackage{amsmath}

\begin{document}

\title{Wave-Growth-Limited Particle Confinement in a Near-Parallel Interplanetary Shock \\ Observed by \textit{Parker Solar Probe}}

\author[0000-0001-6589-4509]{A. Kouloumvakos}
\affiliation{The Johns Hopkins University Applied Physics Laboratory, 11101 Johns Hopkins Road, Laurel, MD 20723, USA.}
\email{Athanasios.Kouloumvakos@jhuapl.edu}

\author[0000-0002-0606-7172]{I.~C.~Jebaraj}
\affiliation{Department of Physics and Astronomy, University of Turku, 20500 Turku, Finland}
\email{immanuel.c.jebaraj@gmail.com}

\author[0000-0001-6360-1987]{M.~A.~Malkov}
\affil{Department of Astronomy and Astrophysics, University of California, San Diego, La Jolla, CA 92093, USA}
\affil{Eureka Scientific, 2452 Delmer St Oakland, CA 94602, USA}
\email{mmalkov@physics.ucsd.edu}

\author[0000-0001-6010-6374]{G. D. Berland}
\affiliation{The Johns Hopkins University Applied Physics Laboratory, 11101 Johns Hopkins Road, Laurel, MD 20723, USA.}
\email{Grant.Berland@jhuapl.edu}

\author[0000-0003-1236-4787]{M.~Gedalin}
\affiliation{Department of Physics, Ben Gurion University of the Negev, Beer-Sheva 84105, Israel}
\email{}

\author[0000-0001-6344-6956]{N. Wijsen}
\affil{Centre for mathematical Plasma Astrophysics, KU Leuven, 3001 Leuven, Belgium}
\email{nicolas.wijsen@gmail.com}

\author[0000-0002-0978-8127]{C. M. S. Cohen}
\affiliation{California Institute of Technology, 1260 E California Blvd, Pasadena, CA 91125, USA.}
\email{cohen@srl.caltech.edu}

\author[0000-0001-6427-1596]{O.~V.~Agapitov}
\affil{Space Sciences Laboratory, University of California, Berkeley, CA 94720, USA}
\email{oleksiy.agapitov@gmail.com}

\author[0000-0001-6160-1158]{D. J. McComas}
\affiliation{Department of Astrophysical Sciences, Princeton University, Princeton, NJ 08544, USA}
\email{dmccomas@princeton.edu}

\author[0000-0003-1960-2119]{D. G. Mitchell}
\affiliation{The Johns Hopkins University Applied Physics Laboratory, 11101 Johns Hopkins Road, Laurel, MD 20723, USA.}
\email{Donald.G.Mitchell@jhuapl.edu}

\author[0000-0002-1349-3663, gname=Edin, sname=Husidic]{E.~Husidic}
\affiliation{Department of Physics and Astronomy, University of Turku, 20500 Turku, Finland}
\email{edin.e.husidic@utu.fi}
            
\author[0009-0003-9858-5720]{S.~Yadav}
\affiliation{Department of Physics and Astronomy, University of Turku, 20500 Turku, Finland}
\email{sunil.s.yadav@utu.fi}

\author[0000-0002-5674-4936]{M. E. Hill}
\affiliation{The Johns Hopkins University Applied Physics Laboratory, 11101 Johns Hopkins Road, Laurel, MD 20723, USA.}
\email{Matthew.Hill@jhuapl.edu}

\author{R. McNutt}
\affiliation{The Johns Hopkins University Applied Physics Laboratory, 11101 Johns Hopkins Road, Laurel, MD 20723, USA.}
\email{ralph.mcnutt@jhuapl.edu}



\begin{abstract}
Collisionless shocks confine the particles they accelerate with the upstream wave field they themselves drive. Streaming particles resonantly amplify Alfv\'en waves, and the waves then confine the same particles by pitch-angle scattering. The rigidity a shock precursor can hold is therefore capped by how fast those waves grow, and that coupling has not previously been constrained in-situ at any interplanetary shock. Here we report Parker Solar Probe measurements of a fast, near-parallel coronal mass ejection-driven shock at 0.24~AU on 2023 March 13 that constrain this wave-particle coupling through the energetic-particle population it accelerates. We measure the spectral folding energy, the boundary between the confined and the upstream-escaping protons, and find that it falls from $\sim$6 to $\sim$1.5~MeV as $E_\mathrm{fold}\propto\Delta t^{\,\alpha}$ with $\alpha\simeq0.34$, where $\Delta t$ is the time remaining before shock arrival. One precursor convective time before arrival this boundary is at $E_\mathrm{c}\simeq 2.46$~MeV, a characteristic confinement energy of a precursor of scale $L_\mathrm{EP}$. Read through the quasi-linear marginal-stability condition, the same evolution implies a resonant wave-growth timescale $\tau_\mathrm{growth}(E)\propto E^{1/\alpha}\simeq E^{2.92}$ that exceeds the precursor convective time above $E_\mathrm{c}$. The post-shock pitch-angle anisotropy reverses sign across the same confinement--escape transition at 2--5~MeV. The rigidity to which a shock precursor can confine particles is therefore set, in-situ, by how quickly the streaming population builds its own confining waves.
\end{abstract}



\section{Introduction} \label{sec:intro}

Intense solar energetic particle (SEP) events are produced by fast coronal mass ejection (CME) driven shocks, and the standard framework for their acceleration is diffusive shock acceleration \citep[DSA;][]{Krymskii1977,Axford1977,Blandford1978,Bell1978}. In its idealized steady-state planar form, DSA predicts a power-law spectrum whose slope is fixed by the shock compression ratio \citep[for reviews see][]{Drury1983,Malkov2001,Lee2012}. However, real CME-driven shocks depart from this idealization in ways that bear directly on observation \citep{Malkov2024}. 

The acceleration is time-dependent. The maximum confined rigidity is set by how quickly the upstream wave field has grown large enough to scatter particles across the shock \citep{Lee1983,Drury1983,Lee2005}. The waves are themselves amplified by the streaming accelerated particles \citep{Skilling1975,Bell1978, Vainio2003, Afanasiev15}, so stronger streaming drives faster wave growth and the enhanced scattering then suppresses the streaming. Particles whose rigidity exceeds the level up to which the waves can be amplified rapidly enough, escape upstream and form the streaming-limit spectrum \citep{Desai2016}, which is a low-rigidity plateau with a folding energy above which the resonant waves have not grown to sufficient amplitude to scatter those particles efficiently \citep{NgReames1994,Vainio2003,ReamesNg2010}. 

Quasi-parallel shocks are the natural laboratory for this coupling. Backstreaming particles drive the resonant ion-beam instability and build an extended foreshock region \citep{Malkov2001}. The near-parallel shock studied here ($\theta_{Bn}\simeq 8^\circ$, the angle between the upstream magnetic field and the shock normal) is an extreme example of this class. At sufficiently strong shocks, additional precursor instabilities include the non-resonant current-driven instability \citep{Bell2004,CaprioliSpitkovsky2014a} and the acoustic precursor instability \citep{DruryFalle86,MalkovDiamond09}, which together with the shock size, strength, and age, set the maximum cosmic-ray energies at astrophysical shocks inferred only remotely \citep{Lagage1983}. In-situ tests remain rare and come mostly from quasi-parallel interplanetary shocks observed near 1~AU \citep[e.g.,][]{Kennel1986} or, more recently, near the Sun \citep[e.g.,][]{McComas2019,Jebaraj2024a,Jebaraj2024b,Wilson2025}. However, the confinement boundary that separates the particles held near the shock from those escaping upstream, and the wave-growth time that sets it, have not been constrained observationally at any interplanetary shock.

On 2023 March 13 at 03:15~UT, a far-side solar eruption produced one of the most extreme CME-driven shocks of the current solar cycle. Parker Solar Probe \citep[PSP;][]{Fox2016} was at 0.24~AU on the inbound of its fifteenth solar encounter, with favorable magnetic connection to the source region and, by inference, to the shock apex. PSP recorded an intense SEP onset promptly after the eruption and crossed the driver shock four hours later. This is the first in-situ encounter with an SEP event of this intensity together with its parent shock inside 0.3~AU. \citet{Jebaraj2024a} reported the shock to be near-parallel ($\theta_{Bn}\simeq 8^\circ$), propagating at $v_{\rm sh}\simeq 2800$~km\,s$^{-1}$ in the spacecraft frame into a magnetized, low-beta upstream ($\beta\simeq 0.2$; \citealt{Jebaraj2026}). The upstream flow speed in the shock rest frame is $u_1\simeq 2400$~km\,s$^{-1}$, and the Alfv\'enic Mach number is $M_A=u_1/v_A\simeq 7.5$ for an Alfv\'en speed $v_A\simeq 320$~km\,s$^{-1}$ in the inner foreshock.\footnote{Throughout this Letter the diffusion calculations are performed in the shock rest frame unless otherwise noted. The Mach number $M_A\simeq 7.5$ uses the refined upstream plasma parameters; \citet{Jebaraj2024a} quoted $9.1$ from a coarser estimate.} The structure of the upstream wave field at this shock is characterized in a companion paper \citep{Jebaraj2026}.

This Letter reports measurements from the PSP Integrated Science Investigation of the Sun \citep[IS$\odot$IS;][]{McComas2016} during this extreme event. We calculate the proton pitch-angle anisotropy from the dispersive SEP onset through the shock crossing, and the energy spectrum from its early upstream-streaming plateau to its at-shock power-law form. We then track the time evolution of the spectral folding energy, which measures the boundary between the confined and the upstream-escaping population, and identify the characteristic confinement energy of the precursor with the value this boundary takes one convective time before the shock arrives. Interpreted through the quasi-linear marginal-stability condition, the same evolution constrains the energy-dependent wave-growth timescale whose crossing with the convective time sets that energy. We emphasise that this confinement energy is not the maximum energy to which the shock accelerates protons. The observed spectrum continues as a power law well above this energy and rolls over well above 10~MeV. The confinement energy marks instead the rigidity above which the energetic particles are no longer held near the shock by the resonant wave field within the precursor.

\begin{figure*}
 \centering
 \includegraphics[width=1\textwidth]{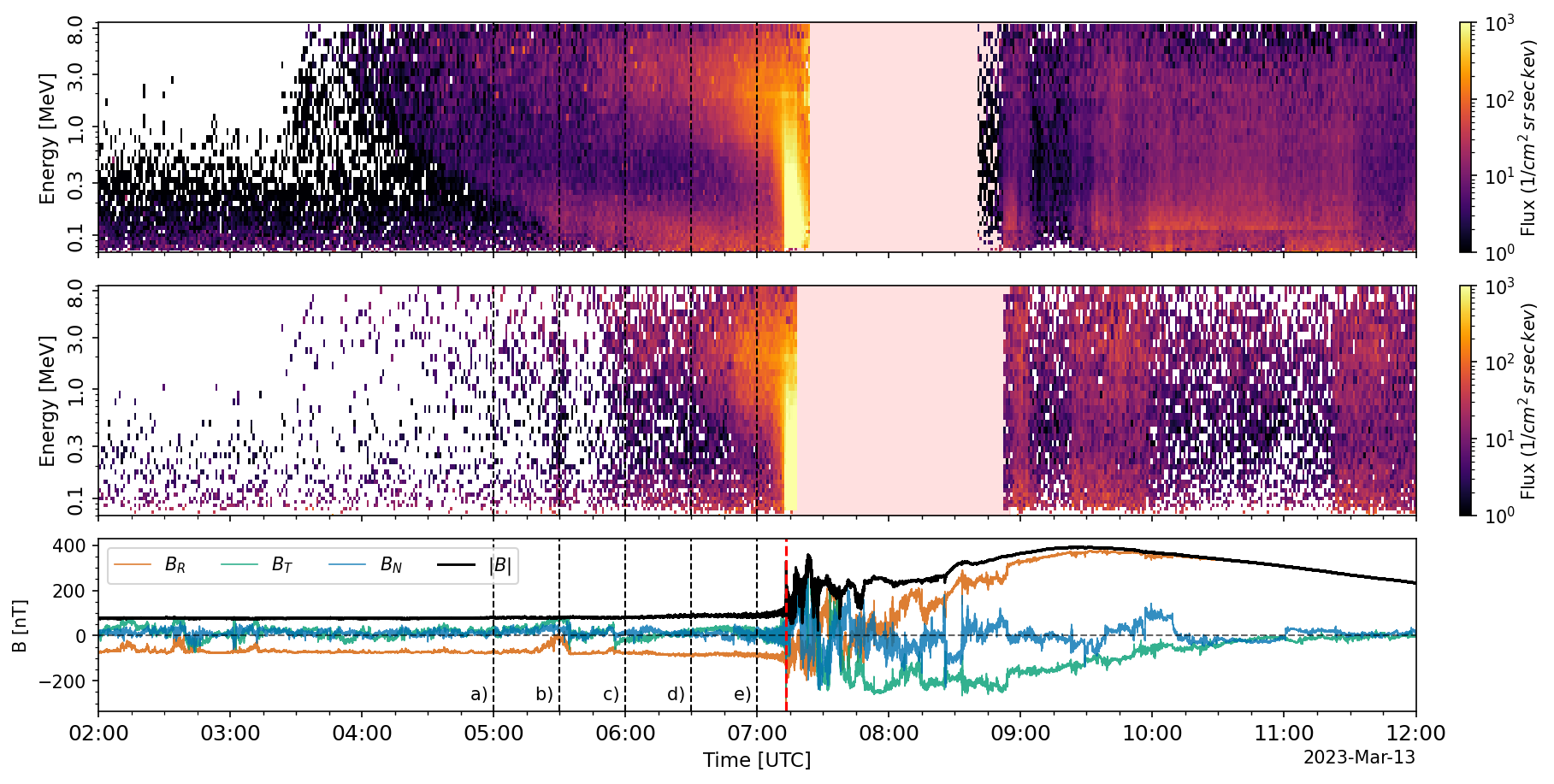}
\caption{Proton energy spectrograms and magnetic-field observations from PSP on 2023 March 13, 02:00--12:00~UT. Top panel: proton differential flux energy spectrograms from PSP/IS$\odot$IS EPI-Lo (ChanP) for the sunward-looking apertures, which sample anti-sunward-streaming particles. Middle panel: same for the anti-sunward-looking apertures, which sample sunward-streaming particles. Bottom panel: PSP/FIELDS magnetic-field components $B_R$ (orange), $B_T$ (teal), $B_N$ (grey), and $|\bm{B}|$ (black) in radial-tangential-normal (RTN) coordinates, with vertical dashed lines marking the five pitch-angle distribution (PAD) snapshot times used in Figure~\ref{fig:SEP_anisotropy_ts}. The shaded band marks the EPI-Lo data gap caused by instrument saturation following the shock crossing.}
 \label{fig:SEP_Observations}
\end{figure*}

\section{Observations} \label{sec:Observations}

\subsection{Event onset and time evolution} \label{sec:Obs_SEPs}

The energetic particle measurements used in this study are from the IS$\odot$IS suite on board PSP. The suite comprises the Energetic Particle Instrument-Low and -High (EPI-Lo and EPI-Hi, respectively) that measure energetic ions from $\sim$20~keV~nucleon$^{-1}$ to $>$100~MeV~nucleon$^{-1}$. In this paper, we primarily analyze measurements of SEP protons from EPI-Lo \citep{McComas_2020, Hill2017}, which is a time-of-flight versus energy instrument with 80 separate apertures distributed between 8 sectors resulting in a hemispherical (2$\pi$-sr) field-of-view. With this extended field-of-view and the amount of distributed apertures, EPI-Lo can provide a high-resolution sampling of the energetic particles' directions. In Appendix~\ref{app:data} we give a detailed description of the SEP instruments and the SEP data used in this study.

Figure~\ref{fig:SEP_Observations} shows the EPI-Lo proton spectrograms separated by looking direction (see Appendix~\ref{app:data} for details) together with the PSP/FIELDS magnetic field \citep{Bale16, Bale_2020} over 02:00--12:00~UT. At onset near 04:00~UT, the sunward-looking apertures (top panel) show a clear velocity-dispersed enhancement across the full EPI-Lo range while the anti-sunward-looking apertures (middle panel) show only a faint signature of this dispersion. The first-arriving particles therefore streamed predominantly anti-sunward, consistent with prompt injection near the Sun onto the field lines connecting the source to PSP. Between $\sim$05:30 and 07:00~UT both spectrograms develop a suppression of fluxes at $\sim$0.2--1~MeV relative to the $>$1~MeV population, most pronounced in the sunward-looking apertures. The directional contrast between the two panels then weakens, indicating progressive isotropization as the shock approaches the spacecraft. The magnetic field is quiet before the shock and becomes highly disturbed within $\sim$15~min of crossing, with strong wave activity upstream (bottom panel).

These are the in-situ signatures of the foreshock developing upstream of the near-parallel shock. Low-energy protons are trapped and further energized by the enhanced wave activity, reducing the upstream-escaping flux below $\sim$1~MeV and producing the sub-MeV spectral flattening seen below (Section~\ref{sec:Obs_Spectrum}). The same scattering, combined with the decreasing distance to the shock, attenuates the strong beam character of the onset (Section~\ref{sec:Obs_Anisotropy}).

\subsection{From injection beam to near-shock isotropy} \label{sec:Obs_Anisotropy}

\begin{figure*}
 \centering
 \includegraphics[width=0.95\textwidth]{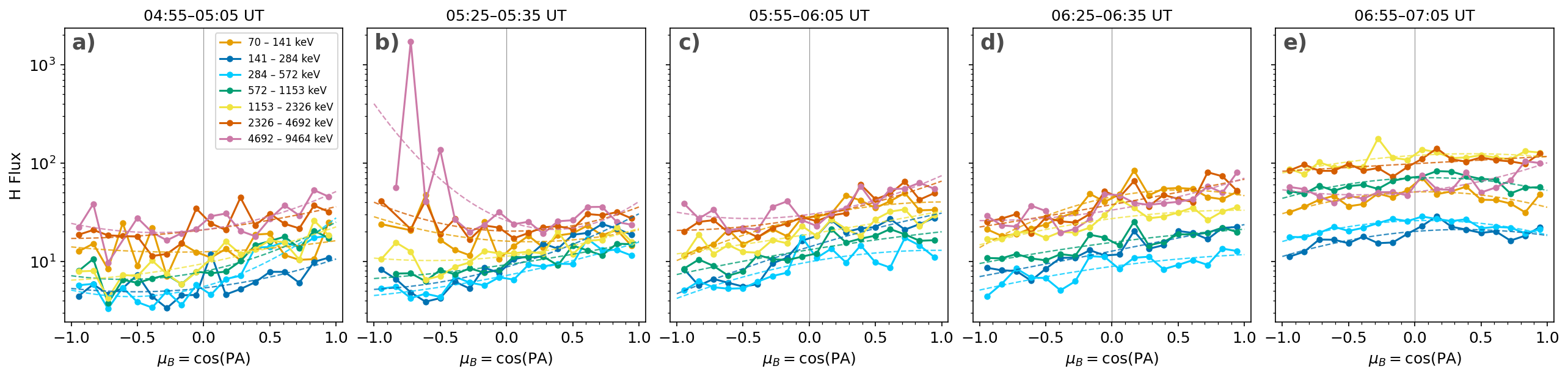}
 \includegraphics[width=1.0\textwidth]{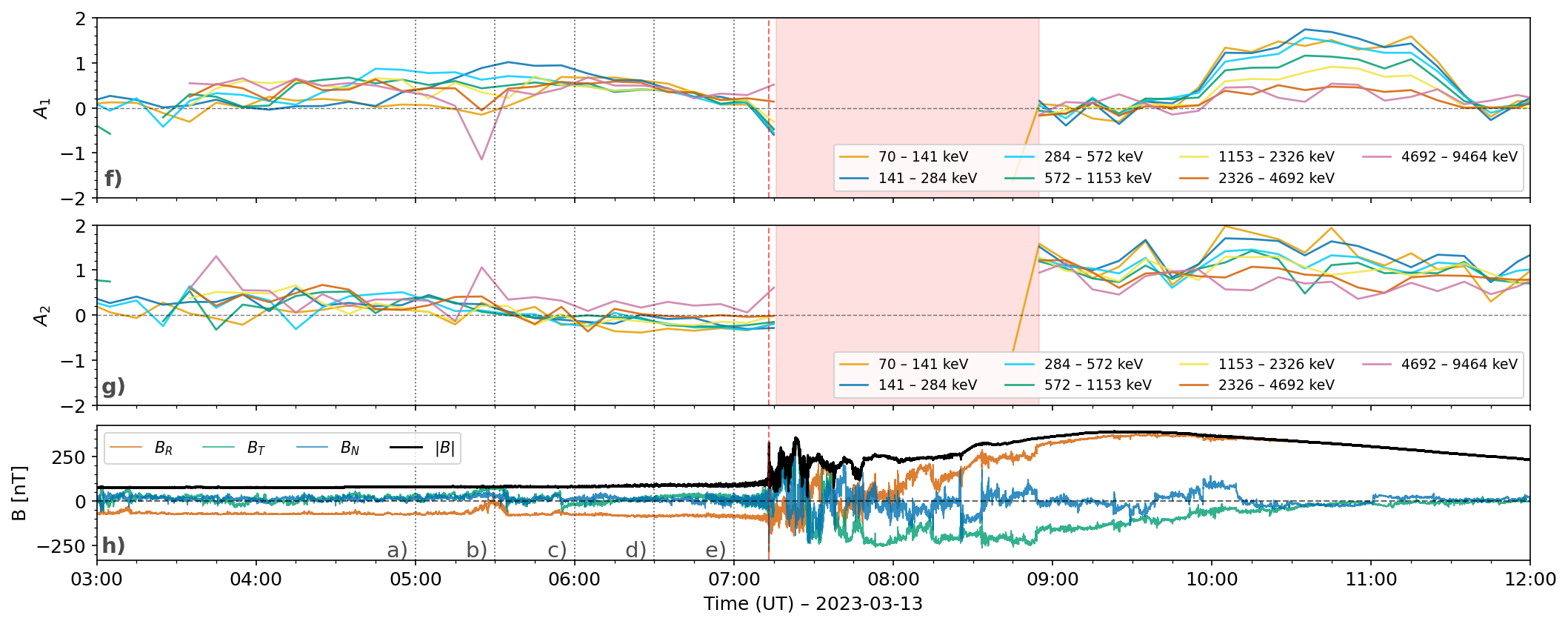}
\caption{Panels (a)--(e): proton PADs in the magnetic-field frame at five selected times on 2023 March 13 and seven EPI-Lo energy ranges, spanning the pre-shock and foreshock evolution (time intervals labeled at the top). Positive (negative) $\mu_B$ corresponds to particles moving purely anti-sunward (sunward) along the local magnetic field ($\mu_B = 0$ is perpendicular to $\bm{B}$). Dashed curves show the best-fit Legendre expansion for each energy. Panels (f) and (g): ten-minute time evolution of the proton pitch-angle anisotropy from 03:00 to 12:00~UT on 2023 March 13. Panel (f) shows the first-order anisotropy $A_1$ (net streaming, with $A_1 > 0$ anti-sunward and $A_1 < 0$ sunward) and panel (g) the second-order anisotropy $A_2$ ($A_2 > 0$ bidirectional along $\pm\bm{B}$, $A_2 < 0$ pancake perpendicular to $\bm{B}$). Panel (h): PSP/FIELDS magnetic-field components in RTN coordinates as in Figure~\ref{fig:SEP_Observations}. The five vertical lines correspond to the five PAD snapshot times.}
 \label{fig:SEP_anisotropy_ts}
\end{figure*}

\begin{figure*}
 \centering
 \includegraphics[width=0.95\textwidth]{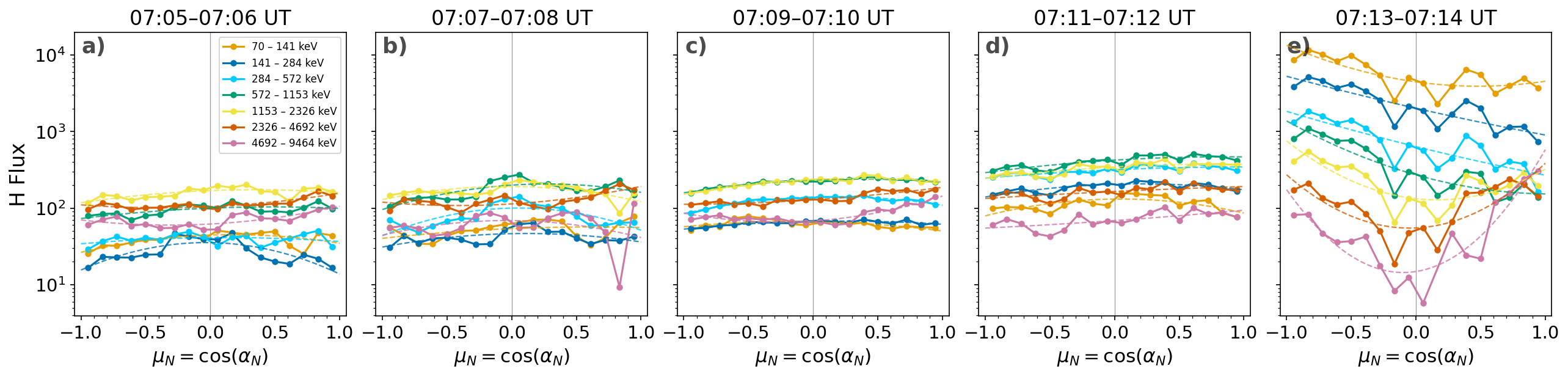}
 \includegraphics[width=1.0\textwidth]{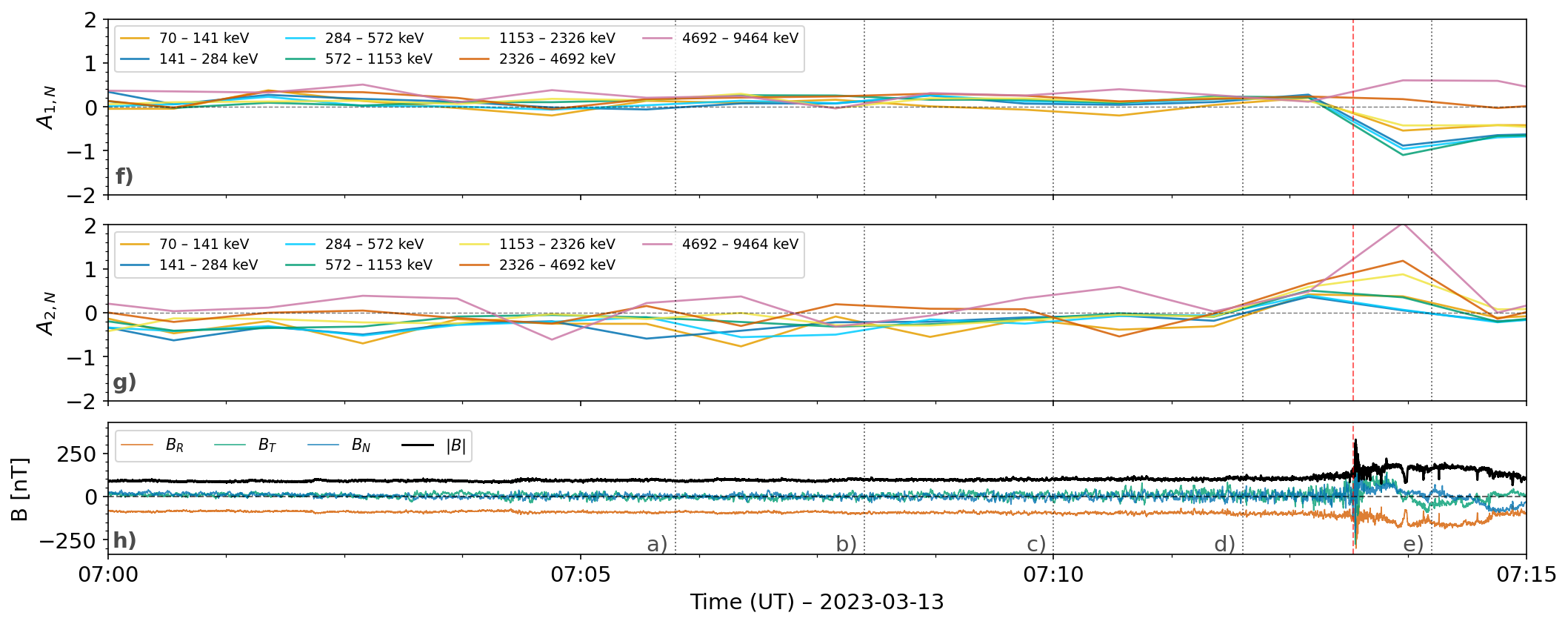}
\caption{As Figure~\ref{fig:SEP_anisotropy_ts}, but for PADs and anisotropy coefficients $A_{1,N}$ and $A_{2,N}$ in the shock-normal frame at one-minute resolution near shock encounter. The shock normal $\hat{N}$ is taken as fixed and pointing from downstream into upstream (anti-sunward in this event), so $\mu_N=+1$ always corresponds to particles moving anti-sunward and $A_{1,N}>0$ denotes anti-sunward streaming. This is motion away from the shock when PSP is upstream, and motion across the shock into the upstream when PSP is downstream.}
 \label{fig:SEP_anisotropy_shock}
\end{figure*}

We bin the proton fluxes from the 80 EPI-Lo apertures into 18 equal-width pitch-angle cosine ($\mu_B$) bins and fit the resulting PADs with a low-order Legendre expansion \citep[e.g.,][]{Sanderson1985}:

\begin{equation}
I(\mu_B) \;=\; \frac{I_0}{2}\bigl[1 + A_1\,\mu_B + A_2\,P_2(\mu_B)\bigr],
 \label{eq:legendre}
\end{equation}

\noindent where $P_2(\mu_B)=\tfrac{1}{2}(3\mu_B^2-1)$, $I(\mu_B)$ is the directional intensity and $I_0$ the omnidirectional intensity, $A_1$ is the first-order (streaming or dipole) anisotropy, and $A_2$ the second-order (bidirectional or quadrupole) component. We adopt $\mu_B=+1$($-1$) as motion along $\bm{B}$ purely anti-sunward(sunward), so $A_1>0$ denotes net anti-sunward streaming. Near the shock we project the same expansion onto the shock-normal direction $\hat{N}$, pointing from downstream into upstream (anti-sunward in this event). The corresponding coefficient is $A_{1,N}$, with $A_{1,N}>0$ denoting motion in the $+\hat{N}$ direction, which is away from the shock when PSP is upstream and across the shock into the upstream once PSP is downstream.

Figure~\ref{fig:SEP_anisotropy_ts}(a)--(e) shows PAD snapshots at five representative pre-shock times. At 04:55--05:05~UT (panel a), a weak but discernible $\mu_B>0$ excess appears in several channels, the early anti-sunward streaming of SEPs along the interplanetary field.

At 05:25--05:35~UT (panel b), the 4692--9464~keV channel develops a sharp field-aligned sunward beam, with flux rising steeply toward $\mu_B=-1$ and a weaker sunward enhancement at 2326--4692~keV. Channels below 2~MeV continue to stream anti-sunward. The episode coincides with a transient magnetic-field discontinuity in panel~(h), where $|\bm{B}|$ drops by tens of nT over a few minutes and then recovers. The reversal occurs only above $\sim$2~MeV. This is a rigidity-selective signature of partial reflection at the structure, affecting only rigidities whose gyroradii are comparable to its scale, while lower-rigidity particles pass through. Such structures are natural sites for non-DSA processing of the precursor population, including drift acceleration and magnetic pumping in the surrounding compressive turbulence \citep{MalkovJebaraj2026}. The episode sits outside the precursor evolution that drives the rest of the analysis, and we do not pursue it here.

At 05:55--06:05~UT and 06:25--06:35~UT (panels c and d), the distributions develop a systematic $\mu_B>0$ excess across most channels, the sustained anti-sunward streaming of particles accelerated by the approaching shock. At 06:55--07:05~UT (panel e), five to ten minutes before shock arrival, all channels are nearly flat, suggesting that strong pitch-angle scattering in the deep foreshock has largely isotropized the population.

Panels~(f) and~(g) show the time evolution of $A_1$ and $A_2$ in all channels. Through the early pre-shock interval ($\sim$03:00--05:00~UT) both anisotropies fluctuate between 0 and 1, with higher-energy channels carrying a weak positive $A_1$ that reflects the mild anti-sunward streaming seen in panel~(a). Around 05:15--05:45~UT, $A_1$ in the 4692--9464~keV channel drops sharply to negative values, a brief episode of sunward streaming of $\gtrsim$2~MeV protons that coincides with the field-aligned beam in panel~(b) and the magnetic discontinuity in panel~(h) as discussed earlier. Lower-energy channels stay strongly positive throughout this episode. From $\sim$06:00 to $\sim$06:30~UT, $A_1$ peaks in all channels, matching the anisotropic PADs of panels~(c) and~(d). As the shock approaches, $A_1$ progressively falls toward zero in all channels, reaching approximate isotropy a few minutes before the shock crossing.

The strongest and most sustained anisotropy develops at and after shock crossing ($\gtrsim$07:13~UT). Figure~\ref{fig:SEP_anisotropy_shock} resolves the sequence at one-minute cadence in the shock-normal frame. From 07:00 to 07:12~UT the distributions are nearly isotropic and both coefficients fluctuate near zero. At the shock crossing, $A_{1,N}$ in the 141--1153~keV channels collapses from weakly positive to strongly negative within a single one-minute bin, then evolves over the next 2--3~min as the downstream distribution reorganizes. With $\hat{N}$ pointing anti-sunward, $A_{1,N}<0$ here corresponds to sunward streaming. This is the same dominant sunward component visible in panel~(h) of Figure~\ref{fig:SEP_Spectrum}, the signature of intense wave-particle scattering organizing the low- and intermediate-rigidity population at and immediately behind the shock. The 2326--4692~keV channel stays near zero. The 4692--9464~keV channel turns clearly positive, with $A_{2,N}$ also significantly positive, consistent with anti-sunward streaming of particles that recross the shock into the upstream plus a bidirectional component from wave-particle interactions near the shock. 

The sign of $A_{1,N}$ near the shock is therefore energy-dependent. Negative at low-to-intermediate rigidities and positive in the highest channels. A net upstream flux from behind the shock requires the downstream diffusive flux to exceed advection, $\kappa_2\,\partial_x f > u_2 f$, with $\kappa_2$ and $u_2$ the downstream diffusion coefficient and flow speed and $x$ measured along $\hat{N}$ into the upstream. Protons above $\sim$4.7~MeV seem to satisfy this condition; the highest-rigidity protons that have been accelerated only recently remain concentrated at the shock and leak back upstream, while lower-rigidity protons, with small $\kappa_2$ and a nearly uniform downstream distribution, are confined.

\subsection{Spectral plateau and at-shock $E^{-1}$} \label{sec:Obs_Spectrum}

\begin{figure*}
 \centering
 \includegraphics[width=1\textwidth]{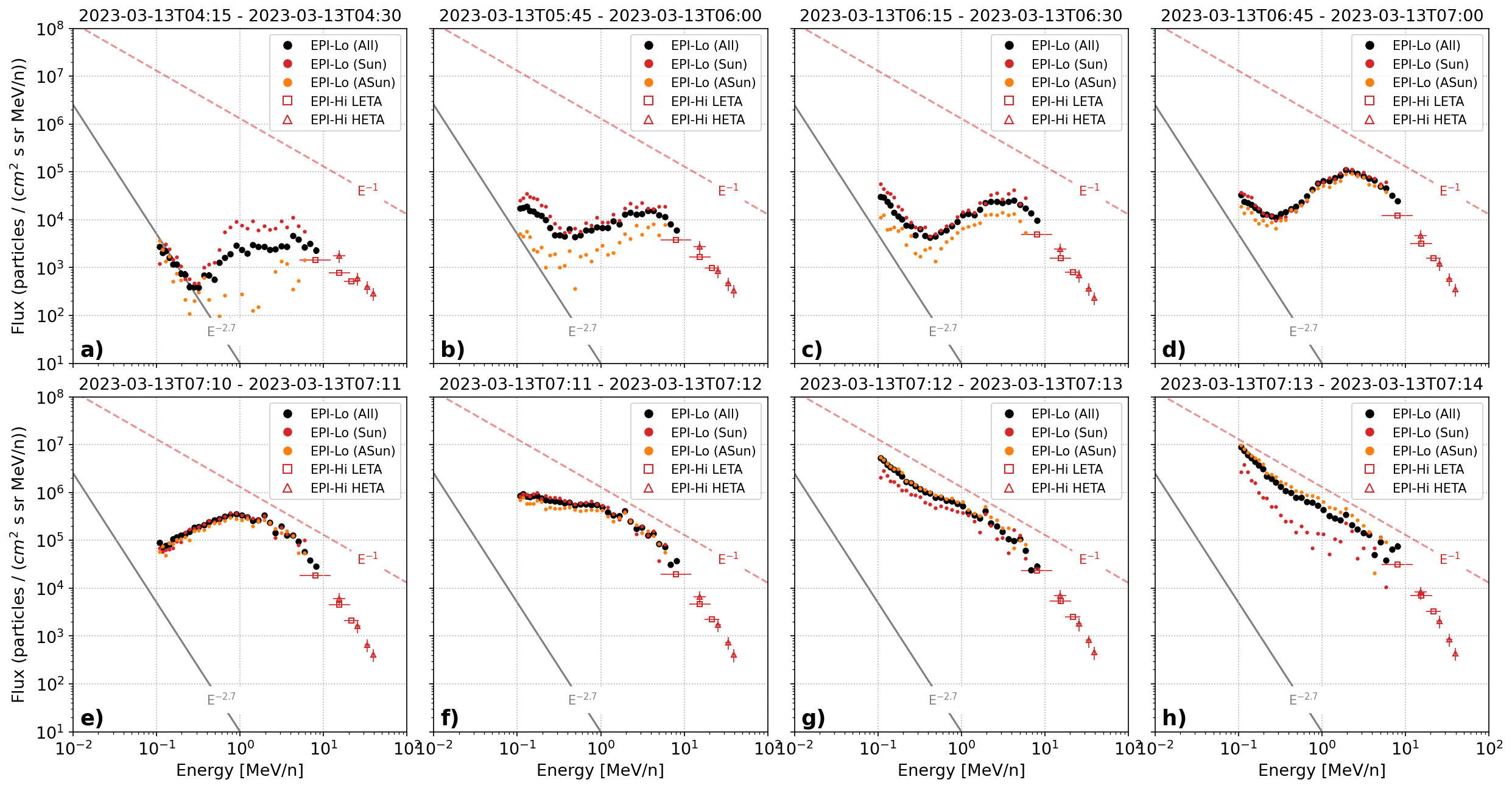}
\caption{Evolution of the proton energy spectrum on 2023 March 13 observed by PSP/IS$\odot$IS EPI-Lo (ChanP, filled symbols) and EPI-Hi (LETA and HETA, open symbols). Black, red and orange symbols correspond to the all-directions, sunward-looking (anti-sunward-streaming) and anti-sunward-looking (sunward-streaming) aperture averages, respectively. Top row (panels a--d): 15-minute averaged spectra at four representative times well before the shock. Bottom row (panels e--h): 1-minute averaged spectra in the four minutes immediately around shock arrival at 07:13~UT (panels e--g upstream, panel h downstream). In every panel the red dashed line marks the asymptotic strong-shock DSA reference $j(E)\propto E^{-1}$; the grey solid line marks the steeper reference power law $j(E)\propto E^{-2.7}$.} \label{fig:SEP_Spectrum}
\end{figure*}

Figure~\ref{fig:SEP_Spectrum} presents the proton energy spectrum at eight representative times. The four upper panels (15-min averages) span the four hours before shock arrival; the four lower panels (1-min averages) span the four minutes immediately around the crossing at 07:13~UT. Black symbols give the omnidirectional EPI-Lo intensities, red and orange the two looking directions (sunward and antisunward, respectively. See Appendix~\ref{app:data} for details). EPI-Hi LETA and HETA extend the spectrum above $\sim$10~MeV. The red dashed line in each panel is the asymptotic strong-shock DSA reference $j(E)\propto E^{-1}$; the grey solid line is the steeper reference power law $j(E)\propto E^{-2.7}$ that fits the background (pre-event) spectrum. To characterize the spectral evolution during the event, we fit a power-law with an exponential rollover (PLER) to each spectrum.

Panel~(a) shows the spectrum at 04:15--04:30~UT, about 15~min after the EPI-Lo onset. The velocity dispersion is still unfolding and the lowest-energy protons have not yet arrived, so the points below $\sim$300~keV reflect the pre-event background. The SEP signal forms a broad plateau from $\sim$0.9 to $\sim$8~MeV, and the EPI-Hi HETA points extend it as a steeper power-law tail with a slope $\gamma \simeq -1.8$ ($j\propto E^{\gamma}$) from the PLER fit.

Panels~(b)--(d) span 05:45--07:00~UT, by which time most of the lower-energy population has arrived. The spectrum departs progressively from the $E^{-1}$ reference. Intensities below $\sim$1~MeV are suppressed by close to an order of magnitude relative to that extrapolation, while the 2--5~MeV decade stays close to it. The rise below $\sim$300~keV is residual pre-event background. The two viewing directions separate most strongly in panels~(b) and~(c), where the intensities in the sunward-looking apertures (red) lie nearly an order of magnitude above the intensities for the anti-sunward-looking apertures (orange) below $\sim$1~MeV. The contrast weakens in panel~(d) as PSP approaches the shock and the distribution isotropizes. This separation is the spectral counterpart of the positive $A_1$ evolution in Section~\ref{sec:Obs_Anisotropy} and identifies the suppressed sub-MeV component as the upstream-escaping streaming population rather than the at-shock distribution. The suppression is sharpest in panel~(d), at 06:45--07:00~UT, where the local maximum at $\sim$2--5~MeV stands roughly an order of magnitude above the suppressed sub-MeV intensities. The PLER fit at this interval gives $\gamma = -2.5$.

Panels~(e)--(h) span 07:10--07:14~UT. The spectrum evolves on minute timescales as the shock approaches. Low-energy intensities rise sharply and the suppression fills in as PSP enters the immediate precursor. At shock crossing (panel h, 07:13--07:14~UT) the low-energy intensities increase by nearly two orders of magnitude. The spectrum becomes a clean power law of slope $\gamma\simeq -1$ across the $\sim$1--10~MeV decade, with a flattening below $\sim$0.5~MeV and a clear rollover in the EPI-Hi HETA channels above $\sim$10--20~MeV consistent with an exponential cutoff at $\sim$15~MeV. This rollover bounds the energy reached over the lifetime of the shock and it is distinct from the spectral folding energy analysed in Section~\ref{sec:Disc_Spectrum} that marks a time-dependent boundary between the confined and the upstream-escaping particle populations. The slope matches the asymptotic strong-shock DSA prediction, $\gamma=-(r+2)/[2(r-1)]=-1$ at compression ratio $r\equiv\rho_2/\rho_1=4$, the non-relativistic energy index following from the strong-shock momentum spectrum $f(p)\propto p^{-3r/(r-1)}$ via $j(E)\propto p^2 f$ \citep{Drury1983}. In panel~(h) the anti-sunward-looking apertures still measure higher intensities than the sunward-looking ones below $\sim$1~MeV, so the distribution remains anisotropic with a dominant sunward-streaming component, the spectral expression of the strongly negative $A_{1,N}$ in the 141--1153~keV channels at shock crossing (Figure~\ref{fig:SEP_anisotropy_shock}).

\section{Discussion \& Conclusions} \label{sec:Discussion}

\begin{figure*}[ht]
 \centering
 \includegraphics[width=0.95\textwidth]{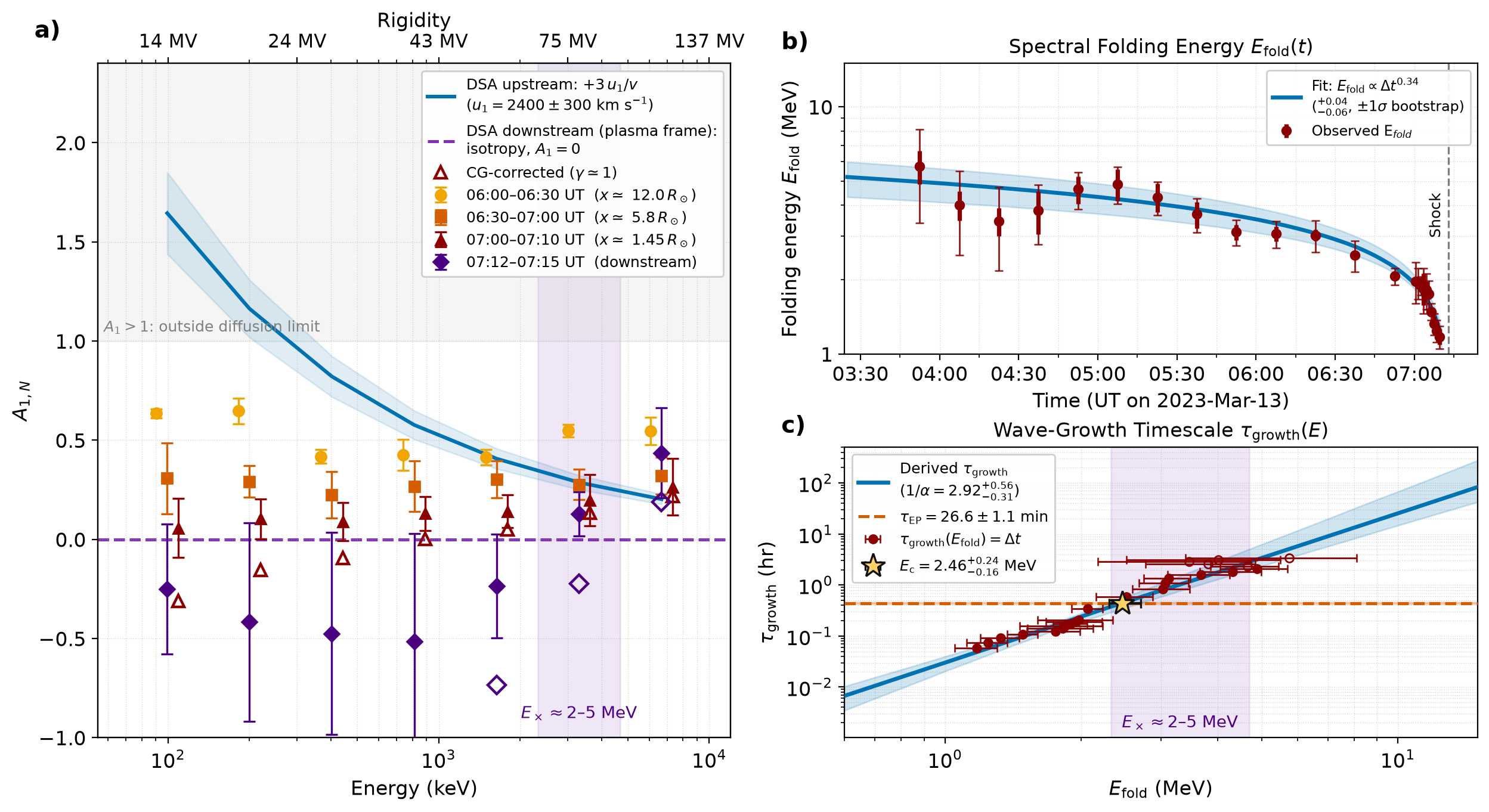}
 \caption{Pitch-angle anisotropy and the inversion to wave-growth time. Panel (a): First-order anisotropy $A_{1,N}$ versus proton energy, with rigidity on the top axis. Coloured filled markers show the observed $A_{1,N}$ from EPI-Lo in three upstream windows and one downstream (see legend). Open markers show the near-shock upstream and downstream anisotropies after the Compton--Getting correction to the plasma frame. The solid blue curve is the upstream DSA ceiling $A_1^{\rm DSA}=3u_1/v$ and the dashed purple line at $A_{1,N}=0$ is the downstream steady-state DSA reference (isotropy in the plasma frame). The grey region ($A_1^{\rm DSA}>1$) marks where the small-anisotropy diffusion approximation breaks down (see Appendix~\ref{app:anisotropy}). The purple shaded band marks the confinement--escape transition $E_x\approx 2$--5~MeV, bracketing the downstream $A_{1,N}$ sign reversal in the spacecraft frame. \textit{Panel (b):} Temporal evolution of the observed spectral folding energy $E_\mathrm{fold}$ (red circles). The inner error bars show the statistical uncertainty and outer bars the statistical and systematic uncertainties combined. The solid blue curve is the power-law fit $E_\mathrm{fold}\propto\Delta t^{\,\alpha}$ and and the shaded band is the 68\% confidence interval from the bootstrap method described in the text. \textit{Panel (c):} Inferred wave-growth timescale $\tau_\mathrm{growth}$ versus folding energy. Red circles show the panel-(b) data re-expressed as $(E_{\mathrm{fold},i},\,\Delta t_i)$ pairs. The solid blue curve is the uniform-$\Gamma_k$ inversion $\tau_\mathrm{growth}(E)\propto E^{1/\alpha}$ and the shaded band its 68\% confidence interval. The dashed orange line marks the precursor convective timescale $\tau_{\rm EP}$ and the star marks the $E_\mathrm{c}$ where the two timescales meet.}
 \label{fig:Theory_Figure}
\end{figure*}

In this study, we have analyzed a particularly strong, quasi-parallel shock observed at a close distance from the Sun. The in-situ Alfv\'enic Mach number $M_A\simeq 7.5$ of this shock sits at the upper end of the distribution modeled by \citet{Kouloumvakos2019} for shocks within $30~R_\odot$ associated with high-energy widespread SEP events. The shock also exhibited significant foreshock wave activity, important for the diffusive shock acceleration of the observed SEPs. The pitch-angle and spectral evolution in Section~\ref{sec:Observations} point to a single picture: a streaming-limited foreshock in which the wave field, driven by the accelerated population to its quasi-linear marginal-stability level, sets both the upstream isotropization and the rigidity at which particles remain confined to the shock \citep{NgReames1994,Vainio2003,VainioLaitinen07,ReamesNg2010}.

Three diagnostics support this picture. The companion analysis of \citet{Jebaraj2026} places the cyclotron-resonant wave power at $k\sim 10^{-3}$~km$^{-1}$, with significant power across a decade in $k$ that resonates with protons from $\sim$0.1 to $\sim$5~MeV. The sub-MeV suppression in Figure~\ref{fig:SEP_Spectrum}(b)--(d) sits inside this resonant band, and the local maximum at $\sim$2--5~MeV marks the lowest rigidity at which the waves can no longer confine particles within the precursor. The at-shock $\gamma\simeq -1$ matches the strong-shock DSA limit ($r\to 4$) expected at $M_A\simeq 7.5$. The next two subsections place the anisotropy and spectral evolution on a quantitative footing.

\subsection{Streaming ceiling and rigidity-dependent scattering} \label{sec:Disc_Anisotropy}

Figure~\ref{fig:Theory_Figure}(a) shows the first-order anisotropy $A_{1,N}$ as a function of proton energy in four time windows. At the same panel, we also show the prediction of linear DSA ($A_1^{\rm DSA}$) as a reference scale against which the wave-mediated rigidity-dependent transport is measured, not as a prediction the data are expected to follow (see details in Appendix~\ref{app:anisotropy}).

In the earliest pre-shock window (06:00--06:30~UT, $x\simeq 12.0\,R_\odot$, yellow symbols), $A_{1,N}$ decreases from $\approx0.6$ at 100--200~keV to $\approx0.4$ at $\sim$0.4--1.5~MeV, before recovering to $\approx0.55$ at 3.3--6.6~MeV. The minimum lies within the resonant band $\sim$0.1--5~MeV identified by \citet{Jebaraj2026}, where the wave power peaks. In the intermediate window (06:30--07:00~UT, $x\simeq 5.8,R_\odot$, orange symbols), the anisotropy decreases relative to the earliest window and becomes more weakly dependent on energy, with $A_{1,N}\approx0.2$--$0.3$ across most of the measured range. The minimum near $\sim$0.5--1~MeV remains broadly consistent with stronger scattering within the resonant band, while the overall reduction in $A_{1,N}$ indicates a progressive isotropization as PSP samples regions closer to the shock. This window therefore represents an intermediate state between the more strongly streaming distribution observed at $12,R_\odot$ and the pronounced rigidity dependence seen immediately upstream of the shock.

In the near-shock window (07:00--07:12~UT, $x\simeq 1.45,R_\odot$, dark-red symbols), a pronounced rigidity dependence emerges. The 100 and 200~keV channels exhibit $A_{1,N}\approx0.04$--$0.10$, indicating a nearly isotropic distribution, consistent with deep isotropization by the resonantly amplified wave field, the condition required to trap and re-accelerate those particles. On the other hand, the anisotropy increases with energy and reaches $A_{1,N}\approx0.27$ at 6.6~MeV. This trend indicates a rigidity dependence of the upstream transport, with substantially stronger pitch-angle scattering at lower rigidities and progressively weaker scattering toward the highest-energy channels. The persistence of a positive anisotropy at 6.6~MeV suggests that the highest-rigidity particles retain a significant streaming component even immediately upstream of the shock, consistent with their longer mean free paths and weaker overall scattering.

This rigidity dependence follows naturally if the wave field is self-generated rather than a background power law. The more numerous low-energy streaming protons amplify short-wavelength waves most efficiently, so higher-rigidity particles resonate with longer-wavelength waves of modest amplitude while lower-rigidity particles are scattered by the strongly amplified short-wavelength waves, steepening the effective rigidity dependence beyond that of any fixed background spectrum. For resonant scattering in a wave field with spectral index $q$ ($W(k)\propto k^{-q}$), quasi-linear theory gives a scattering rate $\nu \propto \Omega_{ci}\,R^{\,q-1}$ for non-relativistic particles of rigidity $R$, growing with rigidity for $q>1$, whereas in the ultra-relativistic regime $\nu \propto R^{\,q-2}$, which decreases with rigidity for the commonly used $1<q<2$ spectra (e.g., Kolmogorov).

The streaming beam, fed by ion reflection and leakage from the quasi-parallel ramp \citep{Malkov1998,CaprioliSpitkovsky2015}, drives a resonant Alfv\'enic instability whose growth rate is set by the field-aligned streaming flux of the resonant protons, the anisotropic first moment of their distribution, so that an isotropic population gives no net growth \citep[][their Eq.~11]{NgReames1994}. The rate is $\Gamma_k \propto \Omega_{ci}\,\Phi_{\rm str}(>p_{\rm res})/(n_{\rm SW}v_A)$ \citep{KulsrudPearce1969,Bell1978,Lee1983}, with $\Omega_{ci}$ the proton cyclotron frequency, $n_{\rm SW}$ the upstream proton density, and $\Phi_{\rm str}(>p_{\rm res})$ the streaming flux carried by protons above the resonant momentum $p_{\rm res}=m_p\Omega_{ci}/k$ for proton mass $m_p$ (gyroradius larger than $1/k$) and growth requiring a streaming speed above $v_A$.\footnote{The non-resonant Bell mode \citep{Bell2004,CaprioliSpitkovsky2014a} is not engaged here. The inner-foreshock wave inventory of \citet{Jebaraj2026} is dominated by the resonant Alfv\'enic branch over the rigidity range of interest.} The waves scatter the same particles that generate them, reducing the spatial diffusion coefficient $\kappa$ and hence $A_1$, so the system relaxes toward the streaming limit \citep{NgReames1994,Vainio2003,ReamesNg2010}. The steady-state $A_1$ is set by the balance between injection at the shock and wave-mediated isotropization upstream. The rigidity dependence of Figure~\ref{fig:Theory_Figure}(a) is consistent with this balance. The wave field is built by, and acts most strongly on, the low-rigidity population that carries most of the streaming flux.

The downstream window (07:12--07:15~UT, dark purple symbols in Figure~\ref{fig:Theory_Figure}a) samples the immediate post-shock distribution. Because the PADs are constructed in the spacecraft frame, open symbols show the near-shock upstream and downstream anisotropies after the Compton–Getting correction to the plasma frame. In the test-particle steady state the downstream distribution is uniform, so the diffusive current vanishes and the distribution is isotropic in the plasma frame. The downstream DSA reference is $A_{1,N}=0$ (dashed purple line). An observer at rest with respect to the shock would instead measure the convective dipole of this isotropic population, $A_{1,N}\simeq -3u_2/v$ with $u_2=u_1/r\simeq 600$~km\,s$^{-1}$ for $r=4$, directed away from the shock. In the 200~keV--1~MeV range there is a high sunward anisotropy which is a signature of intense wave-particle scattering organizing the population immediately behind the shock. Above $\sim$2--5~MeV, $A_{1,N}$ changes sign and the highest channel is positive, which suggests that protons that satisfy the escape condition can leak back upstream. Because the sign reversal of $A_{1,N}$ is bracketed by adjacent channels, we quote the confinement(escape) transition energy, $E_x$, as a band in Figure~\ref{fig:Theory_Figure} (purple shaded region). This band brackets the reversal in the spacecraft frame. The escape condition applies across a boundary at rest in the shock frame, where the downstream flow adds a convective dipole and displaces the reversal to between the 3.3 and 6.6~MeV channels, shown by the open symbols. Escape then requires $A_{1,N}>2(1-\gamma)v_{\rm sh}/v$, the dipole of a population isotropic in the shock frame, which only the 4692--9464~keV channel satisfies. The downstream turbulence is compressionally amplified, so $\kappa_2<\kappa_1$ and confinement behind the shock extends to higher rigidity than the upstream crossover.

\subsection{Wave-growth time and the streaming-limit boundary} \label{sec:Disc_Spectrum}

The rigidity-dependent scattering of Section~\ref{sec:Disc_Anisotropy} has a time-domain counterpart. As the streaming population builds up the wave amplitude, the rigidity at which the field can confine particles evolves. Panel~(b) of Figure~\ref{fig:Theory_Figure} shows the spectral folding energy $E_\mathrm{fold}(t)$, that is, the instantaneous trapping versus escape boundary, where particles with $E<E_\mathrm{fold}$ are confined and re-accelerated near the shock and particles with $E>E_\mathrm{fold}$ escape upstream and form the free-streaming beam. 

We determine the folding energy, $E_\mathrm{fold}$, of the observed proton energy spectrum from the local peak of the spectrum near the transition of the power-law-like spectrum to the low-energy plateau. To reduce random fluctuation affecting the determination of the local peak we used a smoothing cubic B-spline to fit the energy spectrum near the transition and determine the peak from the root of $d\log F/d\log E = 0$ above 0.5~MeV. We also calculated a statistical error $\sigma_\mathrm{stat}$ obtained by resampling the residuals of the spline fit and repeating the peak finding procedure and a systematic error $\sigma_\mathrm{sys}$ which is the spread of $E_\mathrm{fold}$ over a grid of analysis choices, namely the spline smoothing parameter, the pre-smoothing window, the lower bound of the fitted energy range, and the three spectrum constructions (sunward apertures, all apertures with HET or LET data).

From Figure~\ref{fig:Theory_Figure}~(b) we see that as the shock approaches PSP, $E_\mathrm{fold}$ falls from $\sim$6~MeV at $\sim$03:45~UT, set by the first velocity-dispersed high-energy protons that arrive ahead of the $\sim$04:00~UT bulk onset, to $\sim$1.5~MeV just before shock arrival at 07:13~UT. A power-law fit of these points gives

\begin{equation}
E_\mathrm{fold}\propto\Delta t^{\,\alpha},\qquad \alpha = 0.34^{+0.04}_{-0.06},
\end{equation}

\noindent where $\Delta t = t_\mathrm{shock}-t$ is the time until shock arrival. The uncertainties of the fit are obtained from a bootstrap method that resamples the different grid choices and perturbs each point by its own $\sigma_\mathrm{stat}$. The $E_\mathrm{fold}$ measurements before 05:00~UT are excluded from the fit since there, these points carry a 35--38\% systematic error probably because the velocity dispersion at the lowest fitted energy is still unfolding.

The trapping boundary is set physically by the marginal-stability condition $\int\Gamma_k(t')\,dt' = N$. Here $N\equiv\ln(B^2/\delta B_0^2)$ is the number of e-foldings needed for the resonant wave power to grow from the ambient seed $\delta B_0/B\sim 10^{-2}$--$10^{-1}$ to the marginally trapping amplitude, $N\sim 5$--$10$. The streaming-limited wave field this amplification approaches is the self-consistent upstream state of \citet{Lee1983,NgReames1994}, with the resonant growth rate given by Equation~(11) of the latter. Approximating $\Gamma_k$ as uniform over the precursor, the time to amplify the resonant waves from the seed to the trapping threshold is $\tau_\mathrm{growth}(E_\mathrm{fold})=N/\Gamma_k$, which the fold condition sets equal, by construction, to the time remaining before the shock arrives, $\simeq\Delta t$, which is the inflow transit over a length equal to the distance of the sampled parcel from the shock (Appendix~\ref{app:inversion}). Inverting the empirical fit gives

\begin{equation}
\tau_\mathrm{growth}(E)\;\propto\;E^{1/\alpha} \simeq E^{2.9}.
\label{eq:tau_growth}
\end{equation}

The slope $1/\alpha$ is fixed by the empirical $E_\mathrm{fold}(\Delta t)$ scaling and is independent of $N$. The absolute level of $\tau_\mathrm{growth}(E)$ depends on the upstream profile of the population that drives the resonant waves and is the least constrained part of the curve (Appendix~\ref{app:inversion}). Also, the measured slope constrains the drive that builds the wave field. With $\Gamma_k\propto\Phi_{\rm str}(>p_{\rm res})$ (Section~\ref{sec:Disc_Anisotropy}), $\tau_\mathrm{growth}\propto E^{1/\alpha}$ requires the resonant streaming flux to fall as $\Phi_{\rm str}(>E_\mathrm{fold})\propto E_\mathrm{fold}^{-1/\alpha}$ over the developing precursor (Appendix~\ref{app:inversion}).

Panel~(c) of Figure~\ref{fig:Theory_Figure} shows $\tau_\mathrm{growth}(E)$ with the data points $(E_{\mathrm{fold},i},\Delta t_i)$ and the precursor convective timescale $\tau_{\rm EP} \simeq 26.6\pm1.1$~min (the quoted uncertainty is the $1\sigma$ statistical error of that fit). This is the e-folding time of the upstream energetic-particle pressure determined directly from the precursor properties by \citet[][their Figure~4]{Jebaraj2026} from an exponential fit of $P_{\rm EP}$ versus time over the inner foreshock (06:15--07:13~UT) using the same EPI-Lo measurements used here. This inner foreshock region spans $\simeq 12\,R_\odot$, where the wave field is fully developed, and across it the $P_{\rm EP}$ rises by almost an order of magnitude. The corresponding e-folding spatial scale is $L_{\rm EP} = u_1 \tau_{\rm EP} \simeq 5.5\pm0.7\,R_\odot$, with the quoted uncertainty being dominated by that of $u_1$ ($\pm$300~$km\,s^{-1}$). \citet{Jebaraj2026} also show that the implied diffusion coefficient $\kappa_{\rm eff}=u_1 L_{\rm EP}$ matches the quasi-linear estimate at MeV energies for the measured inner-foreshock amplitude $\delta B/B_0\simeq 0.03$, so the precursor scale and the observed wave amplitude are mutually consistent. In Figure~\ref{fig:Theory_Figure} (c) the two curves cross at $E_\mathrm{c}\sim$2.4~MeV, inside the $E_x$ transition band at 2--5~MeV (purple shaded region in panels a and c). This is a confinement boundary and not the maximum energy reached by the shock since the spectrum extends beyond 10 MeV and rolls over at much higher energy. This crossing is the central result of the Letter. Because the inversion rests on the identity of Equation~(\ref{eq:identity}), the crossing has an equivalent definition that invokes no wave-growth model at all,
\begin{equation}
E_\mathrm{c}\;\equiv\;E_\mathrm{fold}(\Delta t=\tau_{\rm EP}),
\label{eq:Ec_def}
\end{equation}
\noindent namely the folding energy measured one precursor convective time before the shock arrives. Read this way, $E_\mathrm{c}$ is a property of the observed confinement boundary alone, obtained by evaluating the empirical fit of panel~(b) at a single time. It is independent of $N$, of the uniform-$\Gamma_k$ approximation (which affect only the absolute level and the high-rigidity end of $\tau_\mathrm{growth}(E)$), and of whether the boundary is attributed to the growth of the resonant waves or to the energy-dependent confinement scale $\kappa_1(E)/u_1$ that the waves set (Appendix~\ref{app:inversion}). Its only external input is $\tau_{\rm EP}$. The wave-growth reading developed here is the physical interpretation of that boundary, not a second, independent measurement. Below $E_\mathrm{c}$ the resonant waves grow faster than the time the precursor plasma spends upstream, so the sub-MeV protons are trapped and rapidly isotropized. Above $E_\mathrm{c}$ the waves cannot be amplified within the available time, and particles scatter only weakly and stream freely upstream through a wave field whose amplitude is set by the lower-energy population. $E_\mathrm{c}$ therefore marks a confinement--escape boundary of the present $\sim$27-min-deep precursor, set jointly by the wave-growth slope and the convective time, and as noted earlier it should not be confused  with the maximum rigidity to which the shock can accelerate particles. The energy spectrum continues as a clean power law through the EPI-Hi range to a rollover energy $\gg$10~MeV (Section~\ref{sec:Obs_Spectrum}), reflecting the four hours of integrated acceleration that built the high-energy tail that EPI-Hi records even at its highest available energy channels. Hence, $E_\mathrm{c}$ is the boundary above which particles outrun the resonant wave field, while the spectrum rollover bounds the acceleration ceiling reached over the shock lifetime.

Two observables of this streaming-limit state locate the transition consistently. The crossing of $\tau_\mathrm{growth}$ and $\tau_{\rm EP}$ in Figure~\ref{fig:Theory_Figure}(c), built from the time evolution of the folding energy $E_\mathrm{fold}$, lands at $E_\mathrm{c}\simeq 2.46^{+0.24}_{-0.16}$~MeV. The post-shock $A_{1,N}$ sign reversal in Figure~\ref{fig:Theory_Figure}(a), built from the reversal of the post-shock streaming, falls in the $E_x$ band at 2--5~MeV, and above 3~MeV in the shock frame. However, the sign reversal is localized only to within one EPI-Lo channel width, in a single three-minute downstream window (07:12--07:15~UT) ending just before EPI-Lo saturates at 07:16~UT, making it difficult to draw solid conclusions from this observation alone.

Other bounds operate in principle in shocks. The Bohm-limited acceleration time \citep{Lagage1983,Lee2005}, the finite-source-size cutoff of a curved, time-dependent shock \citep{Zank2000}, and adiabatic deceleration in the expanding precursor. For the present event the convective time $\tau_{\rm EP}\simeq 27$~min is shorter than $\tau_\mathrm{growth}$ at the rigidities of interest, so the wave-growth limit dominates. 

The same hierarchy is expected at strong astrophysical shocks wherever this inequality holds, with the source age $t_\mathrm{age}$ replacing $\tau_{\rm EP}$ as the relevant time scale. In a young supernova remnant of age $t_\mathrm{age}$, equating $\tau_\mathrm{growth}(E_{\max})\sim t_\mathrm{age}$ with $\tau_\mathrm{growth}\propto 1/\Phi_{\rm str}(>E)$ ties the maximum accelerated rigidity to the streaming spectrum of that system. This is a wave-growth limit, not an acceleration-time limit. \citet{Lagage1983} bounded $E_{\max}$ by the diffusive cycle time in a pre-existing, at-best-Bohm wave field of the ambient medium, while \citet{Bell2004} argued that the streaming particles amplify the confining field far beyond its ambient value \citep[simulation support in][]{CaprioliSpitkovsky2014a}, and the two set opposing expectations for $E_{\max}$. The reconciliation is the growth-time accounting of \citet{Bell2013} where the amplified field must be built within the source lifetime, $\int\Gamma_k\,dt' \simeq N$ e-foldings, the same marginal-stability criterion inverted here, engaged in this event through the resonant rather than the non-resonant branch. The heliospheric and astrophysical cases share this finite-time confinement. They need not share an absolute energy scale or a common streaming spectrum. For a differential streaming density $dn_{\rm str}/dE\propto E^{-s}$, the exponent relating $\tau_\mathrm{growth}$ to rigidity is $s-3/2$ in the non-relativistic regime, as here, and $s-1$ in the relativistic one. Therefore, in-situ interplanetary events can be worked examples of the relation against which astrophysical estimates can be calibrated.

To summarize, we have presented IS$\odot$IS and FIELDS measurements of the intense SEP event of 2023 March 13 and its parent near-parallel shock, encountered by PSP at 0.24~AU. The proton observations show a streaming-limited foreshock in its development phase: suppressed and progressively isotropized sub-MeV SEPs upstream of the shock, an $E^{-1}$ power law at the shock, and a post-shock anisotropy that reverses sign between $\sim$2--5~MeV. From the time evolution of the spectral folding energy we obtain the first in-situ measurement of the confinement boundary of a shock precursor and of its evolution, $E_\mathrm{fold}\propto\Delta t^{\,\alpha}$ with $\alpha\simeq0.34$. Evaluated one convective time \citep[$\simeq$27~min;][]{Jebaraj2026} before the shock, this boundary gives a characteristic confinement energy $E_\mathrm{c}\simeq 2.5$~MeV, consistent with the observed anisotropy reversal as the shock crosses the spacecraft. Interpreted through the marginal-stability condition, the same evolution yields an energy-resolved estimate of the resonant wave-growth timescale, $\tau_\mathrm{growth}(E)\propto E^{1/\alpha}$, which exceeds the convective time above $E_\mathrm{c}$; the slope of this relation is fixed by the measurement, whereas its absolute level rests on the uniform-$\Gamma_k$ approximation (Appendix~\ref{app:inversion}). Below this energy the accelerated protons amplify their confining waves faster than the precursor is convected into the shock and above it the energetic particles escape upstream and form the free-streaming beam. The rigidity to which a shock precursor can confine particles is therefore set by the rate of self-generated wave growth, constrained here for the first time by direct SEP measurements.

\begin{acknowledgments}

We sincerely thank the Referee for his critical review and comments.
This work was supported as a part of the IS$\odot$IS instrument effort on NASA’s Parker Solar Probe mission (NNN06AA01C).
A.K. also acknowledges financial support from NASA’s LWS grant 80NSSC25K0130.
I.C.J. and E.H were supported by the Research Council of Finland (X-Scale, grant No.~371569).
I.C.J. also was supported by FWO grant No.~1295826N. 
I.C.J acknowledges support from ISSI's ``Visiting Scientist Program''.
M.M. was supported by NASA ATP 80NSSC24K0774 and Fermi 80NSSC25K7346 grants.
Parker Solar Probe was designed, built, and is now operated by the Johns Hopkins Applied Physics Laboratory(JHU/APL) as part of NASA’s Living with a Star(LWS) program (contract NNN06AA01C). Support from the LWS management and technical team has played a critical role in the success of the Parker Solar Probe mission. The IS$\odot$IS \citep{McComas_2020} and FIELDS \citep{Bale_2020} data used in this study are available to the community at the NASA Space Physics Data Facility (\url{https://spdf.gsfc.nasa.gov/}).

\end{acknowledgments}

\appendix

\section{SEP Instrument Description and Data}\label{app:data}

The IS$\odot$IS instrument suite \citep{McComas2016} on board the PSP mission, is comprised of the Energetic Particle Instrument-Low and -High (EPI-Lo and EPI-Hi, respectively) that measure energetic ions from $\sim$20~keV~nucleon$^{-1}$ to $>$100~MeV~nucleon$^{-1}$. EPI-Lo \citep{Hill2017} is a time-of-flight versus energy instrument with 80 separate apertures distributed between 8 sectors resulting in a hemispherical (2$\pi$-sr) field-of-view. EPI-Hi \citep{McComas2016} is a set of telescopes of stacked silicon detectors, using the standard dE/dx versus residual energy measurement technique. EPI-Hi LET measure protons from $\sim$1--10~MeV, and the instrument consists of the A, B, and C telescopes where the LET-A telescope is directed toward the Sun, LET-B that is pointing 180$^\circ$ from LET-A directed anti-Sun, and LET-C which is orthogonal to LET-A and LET-B, also HET has the A and B telescopes, where the HET-A telescope is directed 20$^\circ$ west of the radial direction pointing toward the Sun, the HET-B telescope is 180$^\circ$ from the HET-A direction pointing antisunward, and HET measures protons $>$10 MeV. For EPI-Lo we used ChanP \citep[e.g., ``H\_Flux\_ChanP'';][]{McComas_2020} which is a triple-coincidence time-of-flight and total energy measurement provided on a 1 minute cadence covering energies in $\sim$0.07--9.5~MeV range. EPI-Lo observed the event continuously from onset to shock arrival at 07:13~UT and saturated from $\sim$07:16~UT until $\sim$08:55~UT. EPI-Hi was nominally operational but the high intensities triggered its dynamic threshold system which limits the measurement of protons so above $\sim$10~MeV we use single pixel data from EPI-Hi LET and HET.

In Figure~\ref{fig:SEP_Observations} the EPI-Lo data from ChanP are averaged over the sunward-looking(anti-sunward-looking) apertures whose fields-of-view most closely coincide with those of LETA(B) and HETA(B). For the spectrogram with the sunward-looking apertures (Figure~\ref{fig:SEP_Observations} top) we use those with numbers 21--23, 25, 27, 30--37, 41, 44, 46 and for the anti-sunward-looking (Figure~\ref{fig:SEP_Observations} middle) the apertures \#0--1, 4, 61--63, 65, 70--75. For Figures~\ref{fig:SEP_anisotropy_ts} and~\ref{fig:SEP_anisotropy_shock}, the pitch-angle cosine of each aperture is taken from the ChanP looking directions relative to the local magnetic field and folded with the interplanetary-field polarity, with $B_R$ interpolated from the PSP/FIELDS MAG data to each ChanP time stamp. The per-aperture fluxes are averaged over the ChanP channels falling within each of the seven energy bands of Figure~\ref{fig:SEP_anisotropy_ts}, binned into 18 equal-width $\mu_B$ bins on $[-1,1]$, and averaged to a ten-minute (Figure~\ref{fig:SEP_anisotropy_ts}) or one-minute (Figure~\ref{fig:SEP_anisotropy_shock}) cadence. The spectra of Figure~\ref{fig:SEP_Spectrum} are built from the same ChanP one-minute product resampled to the noted time interval for each panel. The omnidirectional spectrum averages all 80 apertures over 0.1--9~MeV, and the two directional spectra average the sunward- and anti-sunward-looking aperture subsets listed above over 0.1--6~MeV. Fluxes are time-averaged over each panel window (15~min in the top row, one minute in the bottom row). Above $\sim$1~MeV the spectra are extended with EPI-Hi LET-A ($\sim$1--30~MeV) and HET-A ($\sim$10--90~MeV) single-pixel rates accumulated over the same time windows and converted to differential flux.

\section{Signed pitch-angle spectrograms}

\begin{figure}
 \centering
 \includegraphics[width=0.45\textwidth]{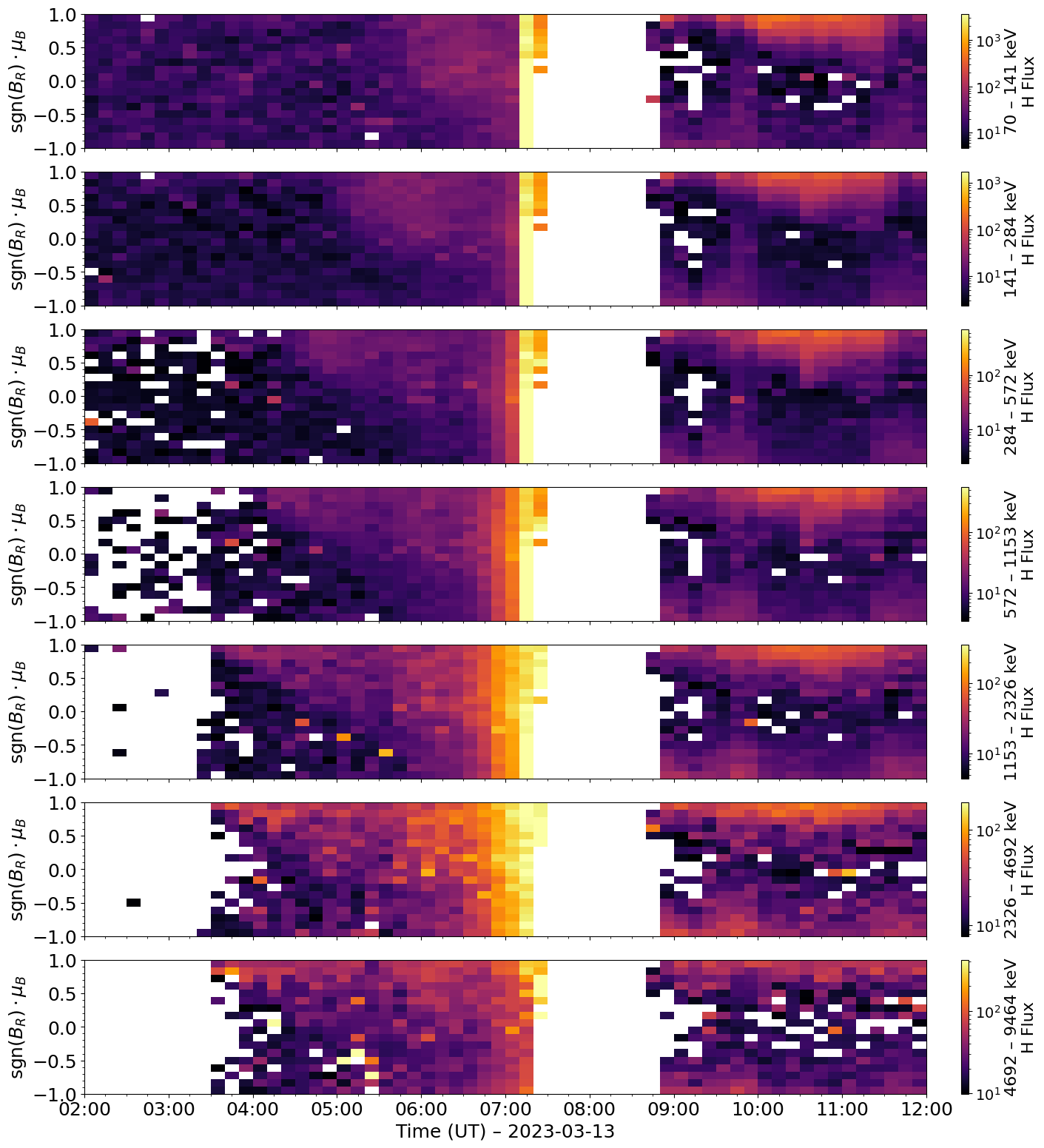}
\caption{Signed pitch-angle cosine $\mathrm{sgn}(B_R)\,\mu_B$ versus time, with $\mu_B = \cos(\mathrm{PA})$, in seven EPI-Lo energy ranges (labeled at right) from 02:00 to 12:00~UT on 2023 March 13. The flux is averaged over the 80 EPI-Lo apertures and binned into 18 equal-width $\mu_B$ bins on $[-1,1]$ at 10-minute cadence. The $\mathrm{sgn}(B_R)$ factor fixes the convention so that positive (negative) values along the vertical axis always correspond to particles moving anti-sunward (sunward) along $\bm{B}$, regardless of the polarity of $B_R$. The color scale is the logarithmic differential flux in $\mathrm{(cm^{2}\,sr\,s\,keV)^{-1}}$.}
 \label{fig:SEP_muB_spectrograms}
\end{figure}

Figure~\ref{fig:SEP_muB_spectrograms} shows the signed $\mathrm{sgn}(B_R)\,\mu_B$ spectrograms for all seven energy ranges, with positive (negative) values on the vertical axis corresponding to particles moving anti-sunward (sunward) along $\bm{B}$. During the early dispersive phase ($\sim$03:30--05:00~UT) SEPs arrive predominantly anti-sunward, with the flux concentrated near $\mathrm{sgn}(B_R)\,\mu_B=+1$. This is the field-aligned beam from the source near the Sun, most pronounced in the highest-energy channels ($>1$~MeV) where the flux contrast across the axis exceeds two orders of magnitude. Between $\sim$05:00 and 06:30~UT the beam fills in toward $\mathrm{sgn}(B_R)\,\mu_B=0$ and below, reflecting progressive isotropization, and by $\sim$06:30~UT the distributions are nearly flat at all energies. In the 10-min bins immediately after shock passage at 07:13~UT and again in the post-gap recovery after $\sim$08:55~UT, the flux maximum shifts to $\mathrm{sgn}(B_R)\,\mu_B\approx -1$ across the intermediate energy bands ($\sim$141--1153~keV) which is a signature of a net sunward-streaming downstream population.

\section{Streaming anisotropy in the diffusion approximation}
\label{app:anisotropy}

In the diffusion approximation the proton phase-space density is expanded in pitch-angle cosine $\mu_N$ about the shock normal and truncated at first order, $F(x,p,\mu_N)=f(x,p)+\mu_N f_1(x,p)$ with $|f_1|\ll f$, where $x$ runs along $\hat{N}$ increasing into the upstream, so $A_1>0$ is anti-sunward streaming away from the shock. The number flux along $\hat{N}$ and the first-order anisotropy are

\begin{equation}
S=\frac{v}{3}\,f_1,\qquad A_{1,N} \equiv\frac{3\int_{-1}^{1}\mu_N F\,d\mu_N}{\int_{-1}^{1}F\,d\mu_N}=\frac{f_1}{f}=\frac{3S}{vf},
\label{eq:flux}
\end{equation}

\noindent with the factor $3=\langle\mu_N^2\rangle^{-1}$. Just upstream of a planar shock the accelerated population streams away from the ramp at the diffusive rate that balances advection, $|S|=u_1 f$ \citep{Drury1983}, giving $A_1^{\rm DSA}=3u_1/v$ (the anti-sunward escaping current). The coefficient defined here is $A_{1,N}$, which for $\theta_{Bn}\simeq 8^\circ$ differs from the field-aligned $A_1$ of by under 1\%, so we drop the subscript throughout. Downstream $|S|=u_2 f$ with $u_2=u_1/r$, and the current reverses to point away from the shock, $A_1=-3u_2/v$. In the scattering-center frame the resonant waves drift outward at $v_A$, so the effective flow is $u_1-v_A$ \citep{Vainio2003}. The expansion holds only where $A_1^{\rm DSA}<1$.

\section{Wave-growth-time inversion}
\label{app:inversion}

Within the marginal-stability model, the spectral folding energy $E_\mathrm{fold}(t)$ that is measured directly from the particles' spectrum, is the rigidity whose resonant waves have just reached the marginally trapping amplitude at time $t$. The marginal-stability condition is $\int\Gamma_k\,dt'=N$, with $N\equiv\ln(B^2/\delta B_0^2)$ the energy-independent e-folding count. Approximating $\Gamma_k$ as uniform over the precursor gives $\Gamma_k\,\tau_\mathrm{growth}=N$, so $\tau_\mathrm{growth}(E_\mathrm{fold})=N/\Gamma_k$. A proton at upstream distance $x=u_1\Delta t$ is confined only if the waves resonant with it have reached the trapping amplitude over the inflow that carried the sampled parcel to $x$, one transit of that distance, and the fold is the rigidity whose growth time equals that transit,

\begin{equation}
\tau_\mathrm{growth}(E_\mathrm{fold})=\Delta t,\qquad \Delta t=t_\mathrm{shock}-t,
\label{eq:identity}
\end{equation}

\noindent an identity independent of $N$. $E_\mathrm{fold}$ therefore falls as the shock approaches, since the remaining time shrinks with the distance.

Streaming particles reach the parcel that PSP samples before the spacecraft does, but the waves it carries are set by the growth integrated along its inflow rather than by how long it has been exposed \citep{Vainio2003}. In steady state the resonant packets are carried into the shock at $u_1-v_A$ and grow at the local rate along the way, so the count at distance $x$ is $\int_x^{\infty}\Gamma_k\,dx'/(u_1-v_A)$ \citep{Bell1978,Lee1983}. Because $\Gamma_k\propto\Phi_{\rm str}$ decays upstream over the precursor scale, this count is the local rate times one transit of that scale, the uniform-$\Gamma_k$ approximation over that transit, and the drive beyond two such lengths adds under 14\%. A drive that is still strengthening in time, concentrates the count further toward the end of the transit. At the fold the precursor scale of the resonant population equals $x$, so the transit is $x/u_1=\Delta t$ to the 15\% by which $u_1$ exceeds $u_1-v_A$, which is Equation~(\ref{eq:identity}). At $E_\mathrm{c}$ the distance $x$ equals the measured scale $L_{\rm EP}$ and the inner foreshock is stationary \citep{Jebaraj2026}, so the transit there is $\tau_{\rm EP}$. Where the resonant waves are driven by higher energies with a longer precursor scale, the transit exceeds $\Delta t$ by a factor that moves the level of $\tau_\mathrm{growth}(E)$ but not $E_\mathrm{c}$.

Inverting the empirical fit $E_\mathrm{fold}\propto\Delta t^{\,\alpha}$ then gives $\tau_\mathrm{growth}\propto E^{1/\alpha}$ (Equation~\ref{eq:tau_growth}). The crossing with the convective time is the most robust output of the inversion. The condition $\tau_\mathrm{growth}=\tau_{\rm EP}$ reduces through Equation~(\ref{eq:identity}) to $\Delta t=\tau_{\rm EP}$, independent of $N$, which is Equation~(\ref{eq:Ec_def}), while the uniform-$\Gamma_k$ approximation affects only the level of the curve through the profile of the drive. The absolute level of $\tau_\mathrm{growth}(E)$ is therefore far less constrained than the energy at which it crosses $\tau_{\rm EP}$. Equation~(\ref{eq:identity}) follows equally if the fold marks the rigidity whose steady precursor scale $\kappa_1(E)/u_1$ equals $x$, so the inversion measures the confinement scale that the growth sets rather than $\Gamma_k$ alone. The growth rate is set by the resonant streaming flux, $\Gamma_k\propto\Phi_{\rm str}(>p_{\rm res})$, so $\tau_\mathrm{growth}\propto1/\Phi_{\rm str}(>E)$ and the integral streaming flux follows the measured slope,

\begin{equation}
\Phi_{\rm str}(>E_\mathrm{fold})\propto E_\mathrm{fold}^{-1/\alpha}.
\label{eq:qmap}
\end{equation}

\noindent Equation~(\ref{eq:qmap}) holds along the locus $E=E_\mathrm{fold}(t)$. Each rigidity being sampled at each time window when it occupies the fold. Since $\Phi_{\rm str}$ rises by close to two orders of magnitude over the precursor, the slope carries this growth together with the rigidity dependence of the drive.

\bibliography{sample702}{}
\bibliographystyle{aasjournalv7.1}



\end{document}